# Thermal stimulated current response in cupric oxide single crystal thin films over a wide temperature range

K. G. Yang, S. X. Wu, F. M. Yu, W. Q. Zhou, Y. J. Wang, M. Meng, G. L. Wang, Y. L. Zhang and S. W. Li

*State Key Laboratory of Optoelectronic Materials and Technologies, School of Physics & School of Materials Science and Engineering, Sun Yat-sen University, Guangzhou 510275, China*

E-mail: stslsw@mail.sysu.edu.cn

**Abstract**

Cupric oxide single crystal thin films were grown by plasma-assisted molecular beam epitaxy. X-ray diffraction, Raman spectrum and *in situ* reflection high-energy electron diffraction show that the thin films are 2×2 reconstructed with an in-plane compression and out-of-plane stretching. Thermal stimulated current measurement indicates that the electric polarization response presents in the special 2D cupric oxide single crystal thin film over a wide temperature range from 130 K to near-room temperature. We infer that the abnormal electric response involves the changing of phase transition temperature induced by structure distortion, the spin frustration and magnetic fluctuation effect of short-range magnetic order, or the combined action of both two factors mentioned above. This work suggests a promising clue for finding new room temperature single phase multiferroics or tuning phase transition temperature.

## 1. Introduction

Multiferroic and magnetoelectric materials, in which ferroelectric and magnetic (ferromagnetic or antiferromagnetic) orders coexist and one of the orders can be modulated by tuning another through magnetoelectric coupling. These materials have been attracting great attention in recent years, because of their interesting physics and promising potential applications [1-3]. The room temperature single phase multiferroic and magnetoelectric materials, such as $BiFeO_3$, however, are extremely rare [3-5]. Cupric oxide (CuO), known as the parent compound of high $T_C$ superconductor, was recently reported as magnetic-induced multiferroics [6-8]. CuO is of high-$T_C$ multiferroic phase between 213 K and 230 K, which is different from other conventional magnetic-induced multiferroics with low-$T_C$ ($T_C$<100 K), such as $RMnO_3$ and $RMn_2O_5$ (R= Tb, Dy, etc) [2, 6-9]. At room temperature, CuO belongs to C2/c space group and has a monoclinic structure (a=4.69 Å, b=3.42 Å, c=5.13 Å; β=99.5°) [10]. CuO undergoes a magnetic transition from paramagnetic phase to incommensurate antiferromagnetic phase (AFM2) with spiral spin order at $T_{N2}$ (~230 K). In addition, a magnetic phase transition from incommensurate (spiral spin) to commensurate (collinear spin) antiferromagnetic phase (AFM1) with spin reorientation occurs at $T_{N1}$ (~213 K). High $T_{N1}$ and $T_{N2}$ could be ascribed to the strong Cu-O-Cu superexchange interaction (J = 60-100 meV) [6, 11-13]. The ferroelectricity of CuO arises from the non-collinear spiral magnetic order which breaks the inversion symmetry and hence induces an electric polarization. The further theory calculation reveals that the ferroelectricity of CuO is intimately related to Dzyaloshinskii-Moriya (DM) interaction and frustrated interaction which enhances the entropy and stabilizes the AFM2 phase [14-18]. Abnormal magnetic and electric properties have been studied in bulk CuO single crystal platelet [6, 8]. Notably, to a certain extent, both experimental and theoretical studies indicate that the structure, properties, and magnetic phase transition temperature of CuO can be effectively modulated under high pressure environment [19-22]. These studies indicate that the pressure induced structure variation and the corresponding change of Cu-O-Cu angle along $[10\bar{1}]$ direction might be responsible for the properties modulation. Inducing a structure distortion or strain in the thin film through substrate stress is one of the most effective ways to achieve a structure modulation.[5] The study on electric properties of CuO single crystal thin film, however, is still absent. In this work, we investigate the single crystal CuO thin films which were epitaxially grown on MgO(111) substrate by plasma-assisted molecular beam epitaxy (PAMBE). The films show abnormal electric polarization response in a wide temperature range from 130 K to near-room temperature, which may imply that the modulation of induced ferroelectric phase could achieve in distorted 2D CuO thin films.

## 2. Experiment

The CuO thin films were grown on MgO(111) substrate (MTI Corporation) using an Omicron customized multiprobe PAMBE system, similar to our previous report [23]. The substrate was cleaned by successive ultrasonic cleaning in acetone and alcohol several times, and then washed in ultrapure water prior to fitting on sample holder. The substrate was annealed 20 mins to desorb the attached contaminations and improve its planeness at the anticipated film growth temperature 500 °C. Atomic oxygen plasma was supplied via an Oxford scientific rf-plasma source under an oxygen partial pressure $3.0\times10^{-5}$ mbar in film growth chamber. The Cu source was provided by a standard high-temperature effusion cell. The temperature of the Cu effusion cell was set at 980 °C to sustain a growth rate around 3 Å/min. After the film growth of 90 mins, $O_2$ flux of $3.0 \times 10^{-5}$ mbar remained 20 mins to improve the crystallization of the as-grown thin films. The growth of the film was monitored by *in situ* reflection high-energy electron diffraction (RHEED), the structure analyzed by X-Ray Diffractometer (D-MAX, Cu Kα), and the thickness (~26 nm) measured by profilometer. The chemical composition of the thin film was analyzed by X-ray photoelectron spectroscopy (XPS, ESCALab250), employing an Al Kα monochromatic x-ray source. Raman spectra measurement was carried out using Renishaw Invia Microscopes Raman Spectrometer with a laser wavelength 514.5 nm. With silver electrodes cold sputtered onto the thin film, thermal stimulated current (TSC) measurement was performed using Keithley 4200-SCS semiconductor characterization system with an accuracy of $10^{-16}$ A. Temperature-dependent electric properties are measured in the vacuum chamber of a cryocooler model RDK-101D cold head of Sumitomo Heavy industry (SHI).

## 3. Results and discussion

X-ray diffraction (XRD) measurement was carried out and the XRD spectrum indicates that the CuO(002) thin film is well crystallized single crystal thin film, as shown in Figure 1. The (002) peak of CuO thin film locates at 35.10°, and the corresponding lattice constant *c* is 5.19 Å, which is a little larger (~1%) than that of CuO single crystal platelet or CuO powder [6, 8]. This could be ascribe to the mismatch between CuO(002) and MgO(111) and the induced compressive stress in the thin film, because the ideal lattice area in CuO(002) plane is ~4% larger than that of MgO(111) plane. The proportion of the ideal in-plane compressing is larger than the out-of-plane stretching, which implies the volume of the thin film is slightly compressed. Raman spectrum of the CuO thin film measured at room temperature is shown in Figure 2. There are three Raman active modes in CuO at room temperature, e.g. $A_g$, $B_g^1$, and $B_g^2$, which locate at 298 cm$^{-1}$, 345 cm$^{-1}$, and 632 cm$^{-1}$, respectively [24]. All the three Raman modes $A_g$, $B_g^1$, and $B_g^2$ only involve the vibration of O atoms [25, 26]. The $A_g$ mode reflects the vibration of O atoms almost along *b* axis, while the $B_g^1$ mode and the $B_g^2$ mode mainly involve the vibration of O atoms along *a* axis and *c* axis, respectively. The Raman shift $A_g$ (305 cm$^{-1}$) and $B_g^1$ (351 cm$^{-1}$) of the CuO thin film is higher, but the $B_g^2$ mode (627 cm$^{-1}$) lower, than that of CuO single crystal platelet. It implies that the $A_g$ and $B_g^1$ modes are obviously hardening, and the $B_g^2$ mode obviously softening and broadening. The $B_g^2$ mode, which mainly involves the vibration of O atoms along *c* axis, shows considerable broadening with an obvious trail to the lower waver number from 627 cm$^{-1}$ to 510 cm$^{-1}$. This is coincident with the result of XRD and implies a strong stress relaxation in the CuO thin film. The RHEED patterns, with different incident electron beam azimuths, of MgO(111) substrate taken from $[1\bar{1}0]$ and $[11\bar{2}]$ direction are shown in Figure 3 (a) and (b). And the other two are the corresponding RHEED patterns of CuO thin film taken from [010] and [100] direction, as shown in Figure 3 (c) and (d), respectively. Twofold periodicity of stripes in both [010] and [100] direction was observed after the film growth, which indicates the film surface is 2×2 reconstructed. This may be ascribe to the mismatch between CuO(002) thin film and MgO(111) substrate. Under a strong in-plane compressive stress induced by the substrate, the atoms on the surface of CuO(002) thin film arrange in 2×2 reconstructed order as a compromise. Combining together with XRD and Raman spectra analysis, we can infer that the CuO thin film is in-plane compressed and out-of-plane stretched. To further investigate the valence state of Cu ions in the CuO thin film, XPS test were carried out. The Cu 2p XPS spectra of the CuO thin film are shown in Figure 4 (a) and (b). The satellite peaks in Cu 2p core-level spectra are the unique d$^9$ configuration character of Cu$^{2+}$ in CuO, as shown in the Figure 4 (a). The fitting analysis of Cu$_{3/2}$ spectra, as shown in Figure 4 (b), indicates that the mix valence states exist in the CuO thin film with Cu$^{1+}$ content around 6%. Different from previous report, this CuO thin film is non-ferromagnetic, which may be ascribed to the relative lower content of mix valence states. The existence of the mix valence states, however, is in favor of breaking the long-range magnetic order and yielding uncompensated spins, which may enhance the spin frustration [23].

The magnetism induced electric polarization response could be reflected by thermal stimulated current (TSC) or pyroelectric current [6, 8, 27-29]. And the multiferroic nanoregions in CuO have a memory effect below $T_{N1}$, thus the electric polarization response could be observed in the heating up process of TSC test begin from the temperature below multiferroic phase transition [8, 27]. In this work, temperature-dependent TSC measurement was carried out along [010] direction from 100 K to 320 K, as shown in Figure 5. Unlike the situation in CuO single crystal platelet, the TSC is abnormal in the CuO single crystal thin film, though there is no report about TSC or pyroelectric current test below 200 K as far as we know. The electric polarization response presents in a wide temperature range from 130 K to near-room temperature. A broad peak locates at 155 K, and the TSC curve shows abnormal fluctuation above 198 K. Besides, there is another broad peak over the temperature range from 232 K to 291 K accompanying with obvious fluctuation. The inset in Figure 5 is enlarged version of the curve from 170 K to 230 K. In this temperature range, the curve shows a small stage with fluctuation between 198 K and 232 K. An abrupt increasing, above 232 K, begins going into the broad peak from 232 K to 291 K. It is notable that a weak fluctuation also exists in the low temperature range 170-198 K.

The underlying mechanism of the abnormal electric response over a wide temperature range is elusive. One possible mechanism is the structure distortion could change the superexchange interaction and the magnetic phase transition temperature. The compressing along *b* axis and stretching alone *c* axis can increase the Cu-O-Cu bond angle alone $[10\bar{1}]$ direction which is known as a key parameter to the antiferromagnetic transition temperature $T_N$ [8, 21, 22]. According to previous reports, the non-collinear magnetic phase or multiferroic phase region in CuO single crystal could be expanded when compressed by high pressure [21, 22]. One may expect, therefore, that the electric polarization response in compressed 2D CuO thin film would present in a wider temperature range than in the bulk. So we infer that the electric polarization response in TSC test, at least partially, arises from the magnetism. The second possible mechanism is the electric response of short-range 1D/2D magnetic order could be likely observed in 2D system over a wide temperature range, or the magnetic order is more stable in the 2D CuO thin film system [30-32]. In this 2D CuO thin film, the existence of mix valence states and induced defects would be benefit to breaking the long rang magnetic order and forming short-range magnetic order in a wide temperature range, which may induce magnetic fluctuation and enhance magnetic frustration that tends to form a non-collinear spin order at high temperature. This point is consistent with the fact that the short-range magnetic order still remains at high temperature (>550K) and a broadened $(\frac{1}{2}, 0, -\frac{1}{2})$ peak had been observed even at 290 K in neutron scattering study of 3D CuO single crystal [8, 11, 12, 30-32]. Thus, the electric polarization could appear through DM interaction. The fluctuation in the TSC curve also supports this point because the magnetic-induced electric polarization fluctuation may result from magnetic fluctuation. On the basis of this scene, we could infer that the abnormal TSC response over a wide temperature range is a novel feature of the short-range magnetic order with spin frustration and magnetic fluctuation in the distorted 2D CuO single crystal thin film system. Or both two factors mentioned above play a role in the abnormal TSC response process. The third possibility is other factors play a role, such as the defects induced by non-stoichiometry [8]. To further understand the underlying mechanism, more experimental and theoretical investigation needs to be carried out.

## 4. Conclusions

In summary, CuO(002) single crystal thin films were grown on MgO(111) using PAMBE method. Combining with X-ray diffraction, Raman spectrum and *in situ* reflection high-energy electron diffraction analysis, we find the CuO thin film is 2×2 reconstructed and structure distorted with an in-plane compression and out-of-plane stretching. Temperature-dependent TSC measurement indicates that the electric polarization present in a wide temperature range from 130 K to near-room temperature in the special 2D CuO single crystal thin film system. In addition, the TSC response shows an obvious fluctuation above 198 K. The abnormal electric response may involve the structure distortion and changing of transition temperature, spin frustration and magnetic fluctuation effect of short-range magnetic order, or the combined action of both two mechanisms. This work refers a possible clue to find room temperature single phase multiferroics or phase transition temperature modulation by structure tuning.

**Acknowledgments**

The authors acknowledge the support from the Scientific Research Foundation for Returned Scholars of Ministry of Education of China, Specialized Research Fund for the Doctoral Program of Higher Education of China (Grant No. 20130171110018), National Natural Science Foundation of China (Grants No. 61273310 and 11304399), and Natural Science Foundation of Guangdong Province (Grant No. 2015A030313121).6

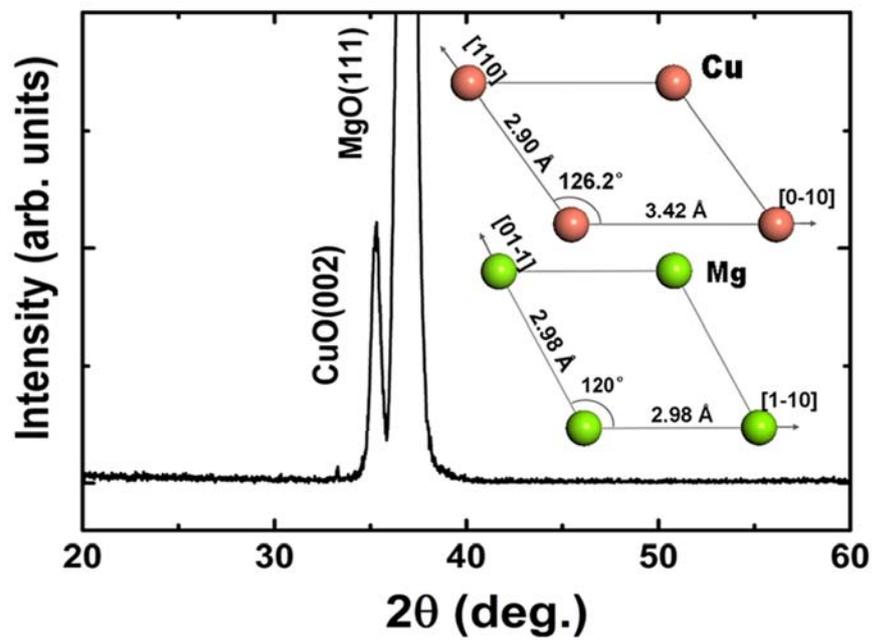

**Figure 1.** XRD spectrum of CuO(002) thin film grown on MgO(111) substrate. The inset is the sketch of Cu lattice in CuO(002) plane and Mg lattice in MgO(111) plane.

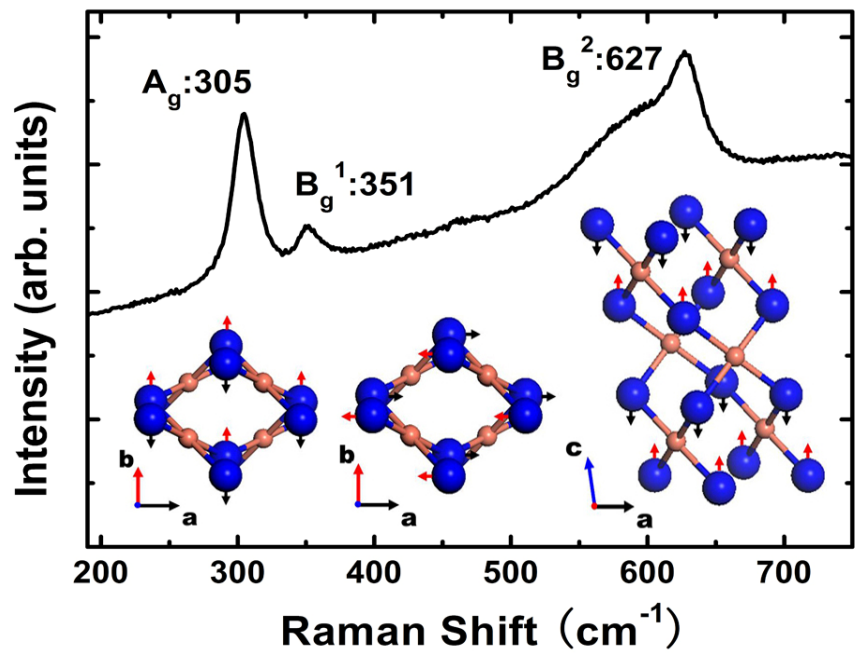

**Figure 2.** Raman spectrum of the CuO thin film and the corresponding O atoms vibration sketches of $A_g$ mode, $B_g^1$ mode, and $B_g^2$ mode.

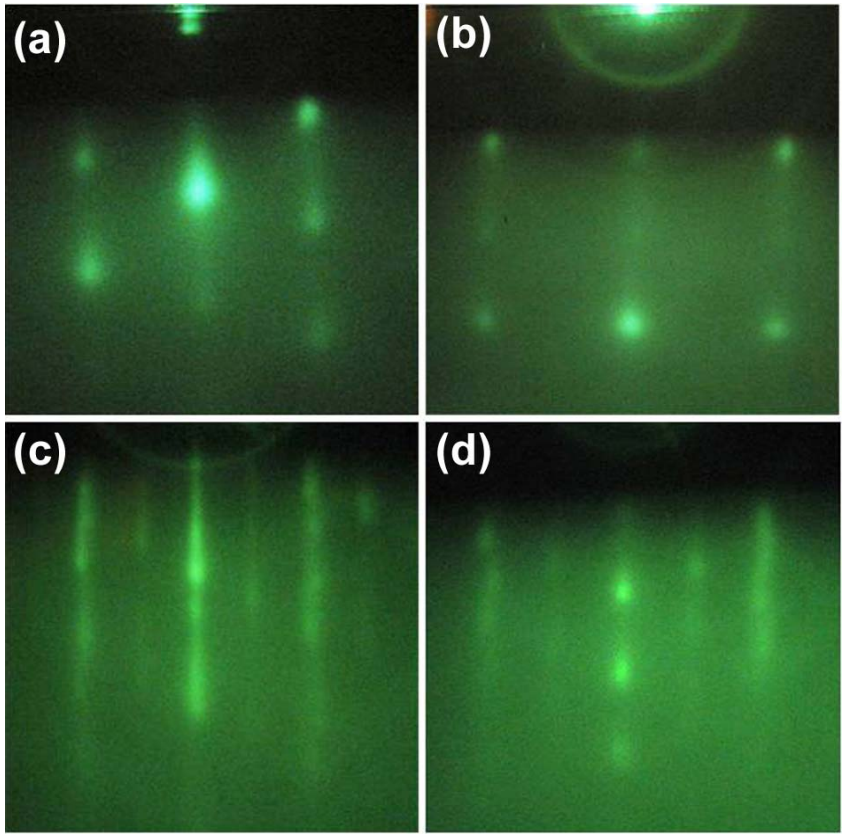

**Figure 3.** RHEED patterns for the bare MgO substrate with different incident electron beam azimuths taken form $[1\bar{1}0]$ and $[11\bar{2}]$ direction (a), (b) before film growth and the corresponding patterns of the CuO thin film taken form [010] and [100] direction (c), (d) after film growth.

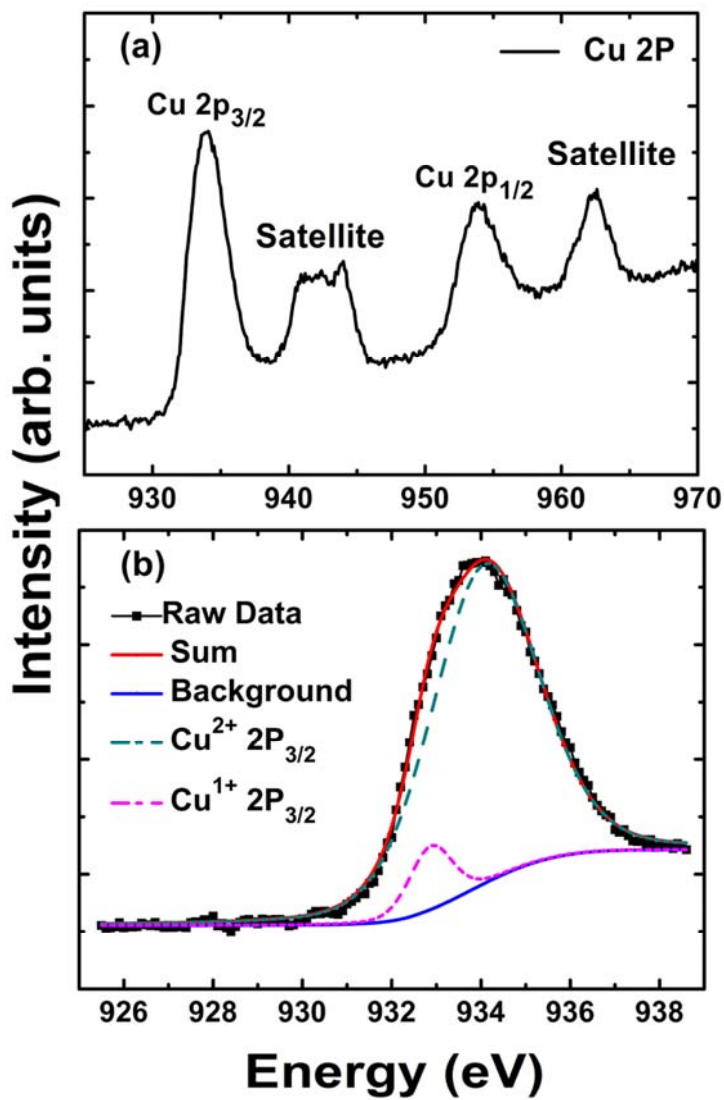

**Figure 4.** The raw data of Cu 2p XPS (a) and the fitting analysis of the Cu $2p_{3/2}$ (b) spectra of the CuO thin film.

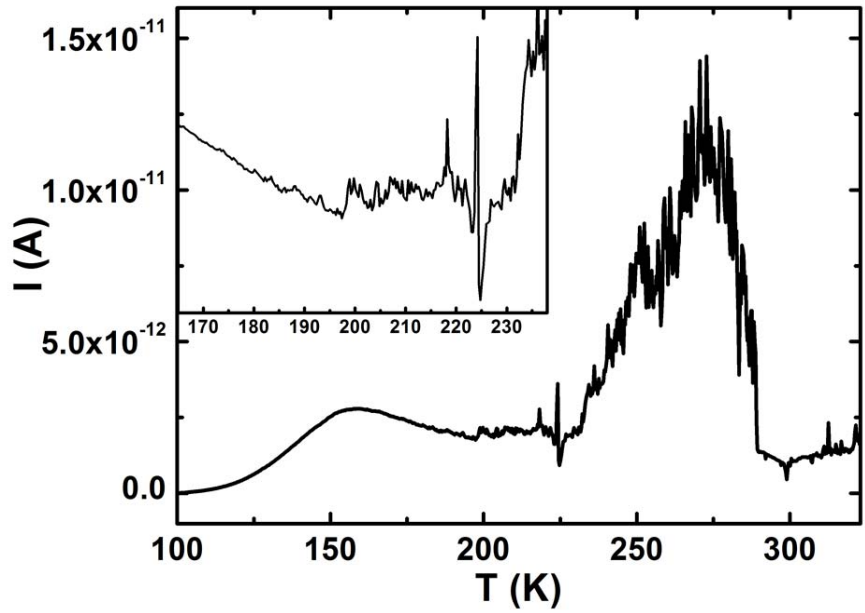

**Figure 5.** The temperature-dependent TSC of the CuO thin film along [010] direction, and the inset is the enlarged version from 170 K to 230 K.